# Exploring the Capabilities and Limitations of Large Language Models for Radiation Oncology Decision Support

Florian Putz[a,b], PD Dr. med., Marlen Haderlein[a,b], PD Dr. med., Sebastian Lettmaier[a,b], Dr. med., Sabine Semrau[a,b], Prof. Dr. med., Rainer Fietkau[a,b], Prof. Dr. med., Yixing Huang[a,b], Dr.-Ing.

[a]Department of Radiation Oncology, Universitätsklinikum Erlangen, Friedrich-Alexander-Universität Erlangen-Nürnberg, 91054 Erlangen, Germany
[b]Comprehensive Cancer Center Erlangen-EMN (CCC ER-EMN), 91054 Erlangen, Germany

Corresponding Author: Yixing Huang (yixing.yh.huang@fau.de)[1]

The advancement of artificial intelligence (AI) has brought forth extraordinary changes across diverse domains, extending its reach to the intricate field of medicine as well. The impressive achievements of AI have set a new bar for what technology can contribute to medicine and, more specifically, to the specialized field of radiation oncology. Large Language Models (LLMs) like GPT-4 represent the apex of AI's progression so far. GPT-3.5, the model's previous iteration, demonstrated notable results by surpassing 50% accuracy across all United States Medical Licensing Exam (USMLE) categories[1]. This provides compelling proof of LLMs' potential to bring transformative change to medical practice.

Thanks to the rapidly evolving integration of LLMs into decision-support tools, a significant transformation is happening across large-scale systems. Like other medical fields, the use of LLMs such as GPT-4 is gaining increasing interest in radiation oncology as well. An attempt to assess GPT-4's performance in radiation oncology was made via a dedicated 100-question examination on the highly specialized topic of radiation oncology physics[2], revealing GPT-4's superiority over other LLMs. GPT-4's performance on a broader field of clinical radiation oncology is further benchmarked by the ACR Radiation Oncology In-Training (TXIT) exam where GPT-4 achieved a high accuracy of 74.57%[3]. Its performance on re-labelling structure names in accordance with the AAPM TG-263 report has also been benchmarked, achieving above 96% accuracies.[4] Such studies shed light on the potential of LLMs in radiation oncology. As interest in the potential and constraints of LLMs in general healthcare applications continues to rise[5], the capabilities and limitations of LLMs in radiation oncology decision support have not yet been fully explored.

**General capabilities:** In the dynamic landscape of radiation oncology, LLMs like GPT-4 are emerging as multifaceted tools. They have the potential to enhance patient self-education, streamline administrative tasks, automate conversations, facilitate research and provide decision support. LLMs also serve as comprehensive reference tools, evident in their potential for Continuing Certification platforms like the ABR's Online Longitudinal Assessment (OLA). For medical training, LLMs have the capability of simulating intricate patient-doctor interactions. Finally, LLMs can efficiently process large amounts of text via in-context-learning (ICL) reaching up to 300 full-text pages per prompt for the upcoming 128k token GPT-4 model. As a large fraction of our collective medical knowledge is encoded in text, LLMs could help us to successfully tap this growing reservoir of medical evidence that otherwise is risking to exceed our conventional information processing capabilities. In this light, the advent of LLMs could signify a transformative shift towards more efficient and informed care. To date, LLMs have already been practically utilized to assist with tasks like creating discharge summaries, patient scheduling and education, medication management, foreign-language translation, and remote monitoring of patient status.

---

[1] This manuscript is officially published as Putz F, Haderlein M, Lettmaier S, Semrau S, Fietkau R, Huang Y. Exploring the Capabilities and Limitations of Large Language Models for Radiation Oncology Decision Support. International Journal of Radiation Oncology, Biology, Physics. 2024 Mar 15;118(4):900-4.

**Objective assessment on complex clinical cases:** Radiation oncology, much like other medical specialties, faces complex clinical scenarios that are often not addressed by existing evidence or guidelines, leaving room for differences of opinion and constructive debates. The Gray Zone cases featured in the Red Journal's Gray Zone case series exemplify such authentic radiooncologic dilemmas. In these intricate scenarios, clinicians could potentially even benefit more from AI decision-support systems than in cases with a well-established gold standard. Benchmarking GPT-4's performance on the Red Journal's Gray Zone cases has been touched[3] (more details in the supplementary material), where the 15 cases of the 2022 Gray Zone collection were preliminarily evaluated from the aspects of correctness, comprehensiveness, and the presence of novel aspects and hallucinations. Nevertheless, the absence of gold standards makes the objective and precise evaluation of LLMs' performance on such authentic, complex radiooncologic cases a formidable challenge, warranting further investigation.

**Current limitations:**

**Knowledge Gaps:** While LLMs have been trained on a broad spectrum of topics and thus encompass a vast expanse of diverse subjects, they are not without their knowledge gaps. In the realm of radiation oncology, GPT-4 has demonstrated inferior performance in specific areas such as radiation measurements[2], gynecologic oncology[3], and brachytherapy[3]. Moreover, GPT-4 also lacks in-depth details of important clinical trials like e.g., the PORTEC-3 trial[3]. Further exploring both the confident and weak zones within LLMs' knowledge base is crucial. This insight not only guides developers in refining these models but also serves as a beacon for users, underscoring the need for rigorous cross-validation.

**Hallucinations:** LLMs can generate incorrect content in a convincing appearance, which is gaining increasing awareness known as "hallucinations". This can be particular dangerous for clinical applications. For a Gray Zone case (Figure S1 in the supplementary material) in our preliminary study, GPT-4 answered in the initial case summary that the patient had "locally recurrent breast cancer with contralateral axillary lymph node involvement", whereas in fact she only had contralateral lymph node metastasis without local recurrence. Moreover, in another case (Figure S2 in the supplementary material), GPT-4 stated that the CATNON and RTOG 9802 trials "demonstrated that adding temozolomide to radiation therapy in patients with grade 2 gliomas and specific molecular markers, such as IDH1 mutation and 1p/19q codeletion, led to improved overall survival". However, RTOG 9802 employed PCV instead of temozolomide chemotherapy and the CATNON trial only included anaplastic glioma. This highlights a challenge that needs to be addressed as AI continues to integrate into healthcare. Therefore, despite the power and sophistication of these models, it remains essential to cross-check all facts provided by LLMs like GPT-4. Both LLM knowledge gaps and hallucinations could be improved by access to external information and traceability, like e.g., implemented the ReACT framework described below.

**Visual understanding:** A critical component of radiooncologic decision-making hinges on the interpretation of medical images, spanning diagnostic and planning CT/MRI scans to dosimetric planning charts. The latest LLMs, such as Bard and GPT-4V(ision), now possess the capability to interpret images as well as generate figures based on text descriptions (GPT-4 with DALLE). While LLMs excel at tasks rooted in natural language processing, they have large limitations when it comes to assimilating visual information for decision support. For example, Bard refuses to process medical images to avoid data privacy and misdiagnosis issues. GPT-4V has the capability of interpreting medical images, but the interpretation can be unreliable, as pointed out by Figure 7 in the GPT-4V system card[6]. Based on our observations, GPT-4V demonstrates its potential in comprehending complex figures in the field of radiation oncology, such as correctly identifying the planning target volume (PTV) curve from a dose volume histogram (DVH) plot (Figure S3 in the supplementary material), describing the linear accelerator (LINAC) system (FigureS4) and the relative dose distribution of different radiation particles in tissue depth (Figure S5). However, it also has the risk of failure for very simple figures, for example, failing to tell the correct number of cells and the correct order of colors in a 4 by 4 grid (Figure S6 (GPT-4V's first attempt) and Figure S7 (GPT-4V's second attempt)). While the second attempt correctly identified the color palette, the spatial arrangement was incorrect, indicating that GPT-4V can distinguish colors but has limitations in mapping these colors accurately to specific spatial locations. This limitation was further underscored in its performance on the Ishihara color test, where it correctly identified only 2 out of 6 numbers (Figure S8), highlighting a potential challenge in interpreting color and spatial relationships, which are both essential for radiation oncology applications.

**Defense of poisoning and attacks:** LLMs are typically fine-tuned using various alignment mechanisms to prevent the generation of inappropriate or obscene content[7,8]. Nevertheless, these models, including renowned ones like GPT-4, Bard, and Claude, remain vulnerable to attacks and poisoning. Intriguingly, even minor attack suffixes appended to queries can provoke these models to produce harmful outputs (e.g., teaching users how to make bombs)[8]. In the realm of radioncologic decision support, there's a palpable risk that a manipulated AI system could disseminate misleading or detrimental information to clinicians and patients. As such, there's an imperative need for the development and deployment of enhanced defense and protection strategies to ensure the safe application of LLMs.

**Data Privacy:** Another pressing concern is data privacy. Most LLMs like GPT-4 are proprietary AI models, and their usage in a clinical setting involves data sharing, which poses significant data security and privacy issues. For instance, OpenAI has faced criticism for allegedly using copyrighted and private data to train GPT-4 without obtaining the necessary consent agreements[9]. Fortunately, open-source LLMs like LLaMA2 with comparable performance to GPT-4 are emerging and can be used locally within hospitals, thereby complying with data privacy regulations. The HuaTuo[10] model exemplifies this approach by fine-tuning LLaMA albeit with specialized Chinese medical knowledge. Similarly, a dedicated model for radiation oncology could be developed leveraging this methodology.

**Overcoming current limitations:**

**Fine-tuning and in-context-learning on specialized medical information:** Although many LLM developers like OpenAI have not publicly disclosed their specific training dataset, it's plausible that their LLMs were trained using a wide array of internet-accessible texts, with specialized medical and radioncologic literature representing only a minor portion. It's highly likely that full-text journal articles and guidelines behind paywalls were not included in the training data. In addition, the wealth of clinical data available in hospitals, e.g., tumor board recommendations, mirrors the potential of specialized literature in enhancing the performance of LLMs. Fine-tuning LLMs on such specialized data or providing LLMs with such data as the input prompt via in-context learning has the high potential to improve their performance on clinical decision making in radiation oncology. An example of this approach is RadOnc-GPT[11], an LLM specifically fine-tuned on an extensive dataset comprising radiooncologic patient records. Further research in this direction is promising to augment the readiness of LLMs for clinical applications.

**Supercharging current LLM models' performance via task-driven agents, access to external information and chain-of-thought reasoning:** As human language has evolved to capture complex thought processes, LLM text prediction can be used to emulate complex reasoning. LLMs are able to decompose complex and multi-step reasoning tasks into a series of intermediate reasoning steps and considerable improve their performance via a chain of thought prompting[12]. Even further improvement is possible by combining chain-of-thought reasoning and LLM self-reflection with autonomous access to external information as has been shown for a ReAct (Reason + Act) method. Intriguingly, ReAct also allows human experts to read the LLM's "chain-of-thought" and check its information sources to follow the model's reasoning. In radiation oncology, a personalized treatment decision relies on a comprehensive consideration of a patient's conditions including cancer type, disease extent, treatment history, molecular pathology and personal priority in concordance with the latest guidelines. Supercharging the capabilities of current LLMs via novel techniques like ReAct could make AI decision support viable. Figure 1 and the supplementary video show an example from a prototypical implementation of the ReAct method for decision support in radiation oncology. In this example GPT-4 provides a sophisticated recommendation for a complex clinical case after a self-determined number of iterations of autonomous guideline tool use, reflection and planning.

**Multimodality integration:** As medical images are pivotal in the realm of clinical practice, the integration of visual and textual information is an emerging research frontier, and many leading LLM providers like OpenAI and Google are actively pursuing advancements in this area. To achieve the goal of multimodality integration, an additional visual model (typically a vision transformer (ViT) or large visual foundation models like the Segment Anything Model) is employed alongside the language model. Due to the deep and wide knowledge embedded in latest LLMs, vision-language models with frozen-weight language models have been widely used, achieving impressive few-shot learning performance on a variety of new tasks[13,14]. The visual information can be integrated to frozen LLMs either at the chat level, as seen in ChatCAD[15] and Visual ChatGPT[16], or at a deeper latent space level, as exemplified in MiniGPT-4[17], Frozen[18], Flamingo[13], and PaLM-E[14]. With the continuous evolution of technology, it is anticipated that more advanced multimodal integration techniques will emerge soon, enhancing the effective utilization of LLMs in clinical radiation oncology.

Due to the inherent limitations of LLM-based decision-support tools, users should exercise caution to avoid overuse of such tools. Well designed, rigorous clinical evaluations are important to identify, in which settings LLMs provide real benefit. Furthermore, a high level of medical expertise is critical for proper use of LLMs for decision-support to assess the value of individual LLM recommendations and decide on optimal patient management.

In conclusion, LLMs like GPT-4 exhibit impressive potential to support decision-making in radiation oncology, but they still have many limitations such as knowledge weak zones, hallucinations, vulnerability to poisoning and attacks, and data privacy issues. Such limitations need to be addressed before broad implementation of LLMs for radiooncologic decision support.

Declaration of Generative AI and AI-assisted technologies in the writing process

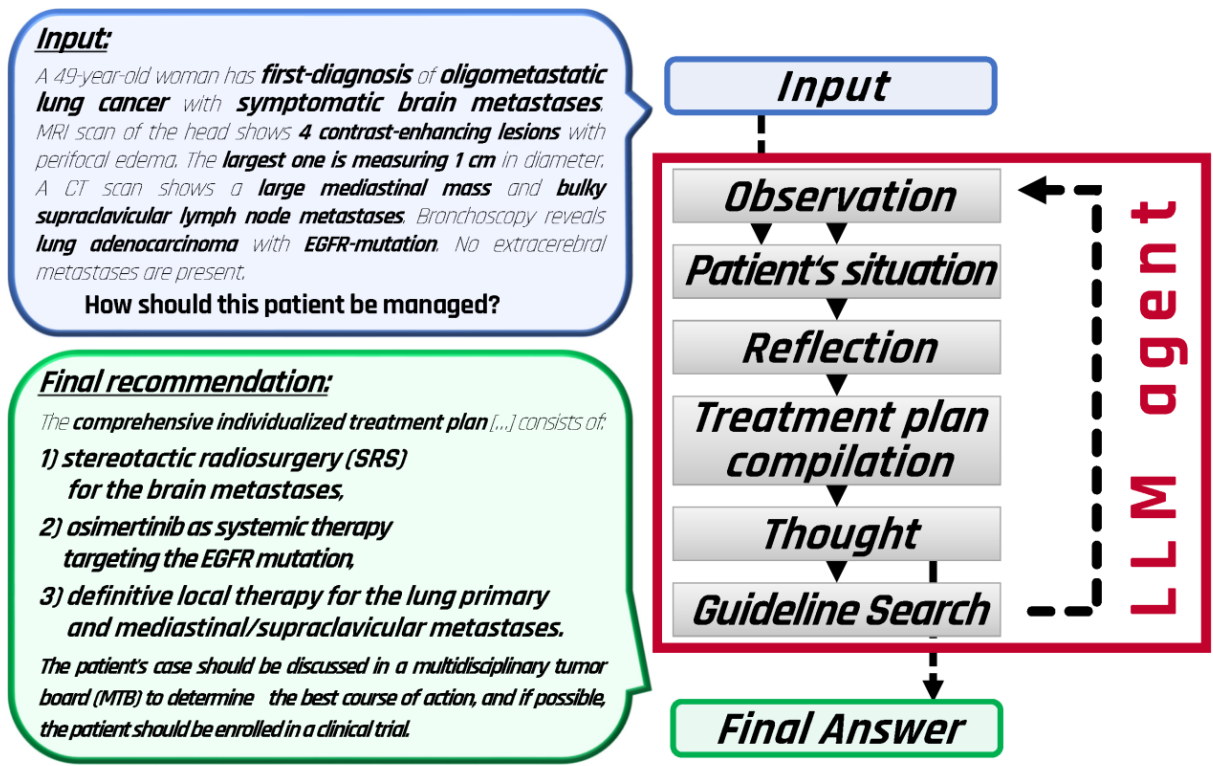

Figure 1: Implementation of a ReAct (Reason + Act) method for decision support in radiation oncology. In this example an LLM acts as an autonomous agent that has access to a semantic guideline search tool, with which it can get access to guideline information by deciding on the specific guideline document and search string. The LLM goes through multiple cycles of guideline search, observation and reflection, while compiling a treatment recommendation. When the LLM has decided that it has gathered enough information and found the optimal solution it can exit the loop and submit its final answer by triggering a "Final Answer" action. In the example shown, GPT-4 provides a sophisticated recommendation for a complex clinical case after a self-determined number of iterations of autonomous guideline tool use, reflection and planning (five guidelines were provided, more details in the supplemental Video).

Supplementary Video download link: https://www.researchgate.net/publication/378610148_1-s20-S0360301623081889-mmc10mp4?_tp=eyJjb250ZXh0Ijp7ImZpcnN0UGFnZSI6ImhvbWUiLCJwYWdlIjoicHVibGljYXRpb24iLCJwcmV2aW91c1BhZ2UiOiJwcm9maWxlIiwicG9zaXRpb24iOiJwYWdlQ29udGVudCJ9fQ

# Supplementary Figures

### Question

You are an expert radiation oncologist from an academic center.
Please give your recommendation on the following case:
"……(text description)"
Question for the expert:
Please give a specific recommendation of your most favored therapeutic approach in this patient case based on the available information and your reasoning for it.

[1] Al-Rashdan A, Cao J. A Viewpoint on isolated contralateral axillary lymph node involvement by breast cancer: regional recurrence or distant metastasis? Int J Radiat Oncol Biol Phys. 2022;113(3):489.

### ChatGPT-4

Based on the available information, the patient has **locally recurrent breast cancer with contralateral axillary lymph node involvement**. Considering the patient's history and pathology findings, my recommendation for the therapeutic approach would include the following:

• Systemic therapy: Given that the patient has hormone receptor-positive (ER/PR+) and HER2-negative disease, initiating endocrine therapy with an aromatase inhibitor (AI) would be my first choice. In addition, considering the prior discontinuation of tamoxifen due to intolerance, AI would be a better option for this patient. Additionally, since the patient has locally advanced and recurrent disease, I would also recommend adding a CDK4/6 inhibitor, such as palbociclib, ribociclib, or abemaciclib, in combination with the AI to improve the efficacy of treatment.
• Radiation therapy: As the patient has not received radiation therapy in the past and given the involvement of contralateral axillary lymph nodes with extranodal extension, I would recommend considering regional nodal irradiation, which includes the chest wall (in the case of a mastectomy) and regional lymph nodes (axillary, supraclavicular, and internal mammary). This will help in reducing the risk of further local-regional recurrence and may improve overall survival.
• Close monitoring: The patient should undergo regular follow-ups with physical examinations, and imaging as needed, to closely monitor her response to the treatment and to detect any signs of disease progression or new metastases.
• Supportive care: **It is essential to address any potential side effects or complications of the treatments mentioned above, such as fatigue, bone health issues, and lymphedema.** This would involve coordinating with a multidisciplinary team of healthcare professionals including medical oncologists, radiation oncologists, surgeons, physical therapists, and other supportive care services.

In summary, my favored therapeutic approach for this patient would be the combination of endocrine therapy with an aromatase inhibitor and a CDK4/6 inhibitor, regional nodal irradiation, close monitoring, and appropriate supportive care. This recommendation is based on the patient's hormone receptor-positive, HER2-negative status, prior treatment history, and the need to address local-regional disease control while minimizing the risk of further recurrence.

### Question

After seeing the other experts' recommendations, are you going to update your own recommendation of your most favored therapeutic approach in this patient case? If so, what is your updated recommendation?

### ChatGPT-4

After reviewing the other five experts' recommendations, I would like to update my initial recommendation to provide a more comprehensive approach for the patient.

Updated recommendation:
• Perform thorough imaging, including bilateral magnetic resonance imaging, to assess for potential primary or recurrent tumors in the residual ipsilateral or contralateral breast tissue.
• Administer right-sided locoregional radiation therapy, targeting the right-sided axillary levels 1 to 3 and supraclavicular nodes. Consider bilateral comprehensive radiation therapy if imaging or further assessment indicates the necessity.
• Initiate systemic therapy, including endocrine therapy with an aromatase inhibitor, a CDK4/6 inhibitor, and possibly chemotherapy to mitigate the risk of distant recurrence and maximize progression-free survival and overall survival.
• Conduct close monitoring and long-term follow-up to detect any signs of recurrence or progression early and provide appropriate supportive care.

This updated recommendation combines essential elements from the other experts' suggestions while balancing the need for aggressive treatment and minimizing potential side effects. It provides a comprehensive and multidisciplinary approach tailored to the patient's needs and overall prognosis.

Figure S1

## Question

You are an expert radiation oncologist from an academic center.
Please give your recommendation on the following case:     [1] Scarpelli D, Jaboin JJ. Exploring the role of resection post-radiation therapy in gliomas.
"……(text description)"                                      *Int J Radiat Oncol Biol Phys.* 2022;113:11.
Question for the expert:
Please give a specific recommendation of your most favored therapeutic approach in this patient case based on the available information and your reasoning for it.

## ChatGPT-4

Based on the available information, my recommendation for this patient would be to consider a multidisciplinary approach that includes a thorough reevaluation of the possibility of maximal safe resection of the residual tumor, followed by continuing adjuvant temozolomide, and potentially the addition of tumor-treating fields (TTF) if deemed appropriate. This recommendation is based on the following reasoning:

• Maximal safe resection: The patient initially underwent partial resection of the tumor, and the treatment team deferred completion resection. However, given the decrease in the expansile T2/FLAIR mass after chemoradiation, it may be worth reevaluating the feasibility of achieving a more complete resection. Maximal safe resection of low-grade gliomas has been associated with improved overall survival and progression-free survival in multiple studies. If the multidisciplinary team, including a neurosurgeon, deems it safe and feasible, this could be a beneficial step for the patient.
• Continuing adjuvant temozolomide: The patient is currently on their 5th cycle of adjuvant temozolomide, which is in line with the treatment recommendations from the CATNON and RTOG 9802 trials. These trials **demonstrated that adding temozolomide to radiation therapy in patients with grade 2 gliomas and specific molecular markers, such as IDH1 mutation and 1p/19q codeletion, led to improved overall survival**. Given the patient's positive response to treatment so far, it would be reasonable to continue with the remaining cycles of adjuvant temozolomide.
• **Consideration of tumor-treating fields (TTF):** TTF is a novel treatment modality that uses low-intensity, alternating electric fields to disrupt cell division in cancer cells. In the EF-14 trial, the addition of TTF to maintenance temozolomide was shown to improve overall survival and progression-free survival in patients with newly diagnosed glioblastoma. While the trial specifically focused on glioblastoma, TTF may still be considered as a potential therapeutic option for patients with grade 2 gliomas, particularly if the multidisciplinary team believes it could be beneficial in this specific case.

In summary, I recommend a thorough reevaluation of the possibility of maximal safe resection of the residual tumor, continuing adjuvant temozolomide, and potentially considering the addition of TTF if deemed appropriate by the multidisciplinary team. The patient's case should be regularly discussed and monitored by a team of experts, including a neurosurgeon, radiation oncologist, and medical oncologist, to ensure the most effective and personalized treatment plan.

## Question

After seeing the other experts' recommendations, are you going to update your own recommendation of your most favored therapeutic approach in this patient case? If so, what is your updated recommendation?

## ChatGPT-4

After reviewing the recommendations of the other three experts, I would update my initial recommendation slightly, while still maintaining a combination of Expert 1's and Expert 2's suggestions:

Updated Recommendation:
• Continue with observation and complete the scheduled maintenance temozolomide, as the patient has favorable molecular markers and has shown a decrease in the tumor burden after chemoradiation.
• Reevaluate the possibility of maximal safe resection of the residual tumor, involving a multidisciplinary team, including a neurosurgeon, to determine if it is safe and feasible to achieve a more complete resection, taking into consideration the patient's good Karnofsky performance status.
• I would no longer emphasize the potential addition of tumor-treating fields (TTF) as a primary recommendation in this case, given the other three experts did not mention it and the focus should remain on the management strategies with more established benefits for this specific patient population. However, TTF could still be considered as an option for discussion within the multidisciplinary team if the standard approaches are not providing the desired results or if further evidence emerges supporting its use in grade 2 gliomas.

This updated recommendation is based on a combination of Expert 1's and Expert 2's recommendations, as it acknowledges the importance of completing the ongoing adjuvant therapy while also considering the potential benefits of maximal safe resection if deemed appropriate by the multidisciplinary team.

Figure S2

**You**

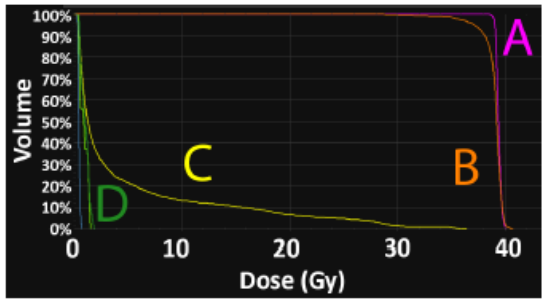

Attached you find a Dose-Volume-Histogram (DVH, % volume of structure / target vs. Dose in Gy) of a patient with multiple myeloma that was treated at a right orbital metastasis with 39 Gy in 3 Gy fractions.

Please derive which structures/targets the curves correspond to just from the trajectory of the curves. First explain how to separate different structures and targets (including GTV vs. PTV) in the DVH based on their curves. Then please go through the lines described: A, B, C and D and derive the underlying structures/targets. Which insights can be derived from the DVH?

**ChatGPT**

The Dose-Volume Histogram (DVH) you've provided is a tool used in radiation therapy to evaluate the distribution of the dose received by the tumor and the surrounding tissues. Each curve on the DVH represents a different structure or target volume, such as the Gross Tumor Volume (GTV), the Planned Target Volume (PTV), and various organs at risk (OARs).

Here's how to interpret the curves:

1. **GTV vs. PTV**: The GTV is the gross palpable or visible/detectable extent and location of the malignant growth. The PTV is a region that typically encompasses the GTV plus a margin to account for microscopic malignant disease around the GTV and for patient motion and treatment setup variations. The GTV should receive the highest dose, shown on a DVH as a steep curve that rises quickly to 100% volume at high doses. The PTV will have a similar shape but might extend to lower doses due to the margin around the GTV which might receive slightly less dose.
2. **OARs**: Organs at risk are typically represented by curves that should ideally receive much lower doses than the GTV or PTV, thus their curves will be to the right and below the curves for the GTV and PTV, showing that a lower percentage of their volume receives high doses.

Now, let's go through the lines described in your DVH:

- **Curve A**: This curve rises sharply to 100% volume at the highest dose, suggesting it represents a target volume receiving the full prescribed dose, possibly the GTV or PTV.
- **Curve B**: This curve is also very steep, but it reaches 100% volume at a slightly lower dose than Curve A, indicating that it could be the PTV, which includes a margin around the GTV.
- **Curve C**: This curve extends over a range of doses, suggesting it is an organ at risk that should receive as little dose as possible, but due to proximity to the target volumes, it inevitably receives some dose.
- **Curve D**: This curve represents a structure receiving a very low dose compared to the other structures, which might be either a non-critical structure or an organ at risk that is well-spared.

From this DVH, one can derive insights regarding the adequacy of tumor coverage and the sparing of healthy tissue:

- The steepness of Curves A and B suggests a high dose is being delivered to the target volumes, which is desirable for treatment effectiveness.
- The dose fall-off between the target volumes and the OARs (Curves C and D) indicates the effectiveness of sparing healthy tissue.
- The specific shapes and separation of the curves can tell us about the precision of the treatment delivery, with sharp curves indicating precise targeting.

However, without specific labels or additional clinical information, these interpretations are somewhat speculative. In practice, each curve would be labeled with the structure it represents, and treatment decisions would be based on detailed knowledge of the anatomy and the treatment plan objectives.

Figure S3

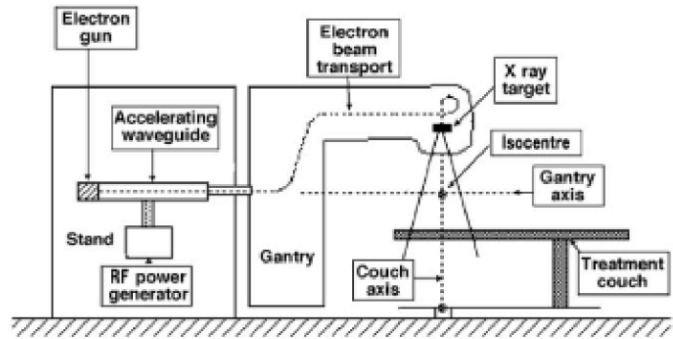

Figure S4

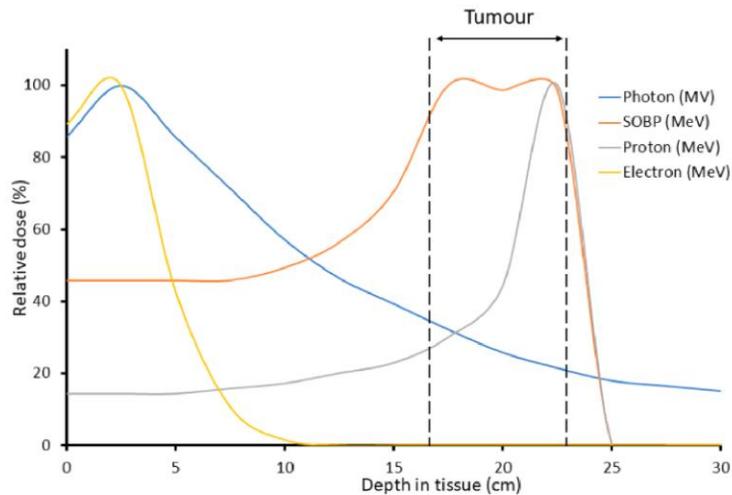

Can you please describe this image?

Certainly!

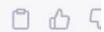

The image depicts a graph showing the relative dose (%) of different types of radiation particles as a function of their depth in tissue (measured in centimeters, cm). The vertical axis is labeled "Relative dose (%)" and ranges from 0% to 100%. The horizontal axis is labeled "Depth in tissue (cm)" and spans from 0 cm to 30 cm.

Four curves are presented on the graph, each representing a different type of radiation:

1. Photon (MV): This curve is blue and shows a gradual increase in dose as depth increases, reaching a peak before gradually decreasing.
2. SOBP (MeV): This curve is orange and rises sharply in the beginning before plateauing and maintaining a consistent dose across a range of depths. Then, it sharply drops off.
3. Proton (MeV): This curve is gray and rises quickly to a peak dose at a specific depth, then sharply declines to zero.
4. Electron (MeV): This curve is yellow and shows a rapid increase to its maximum dose at a shallow depth before rapidly decreasing.

There is also a dashed vertical line labeled "Tumour" that seems to indicate the ideal depth at which a tumor is located. This line helps in comparing how each radiation type delivers its dose relative to the tumor's position.

The graph provides insights into how different radiation types deliver doses at various depths, which is critical in radiation therapy for treating tumors effectively while minimizing damage to surrounding healthy tissue.

Figure S5

**You**

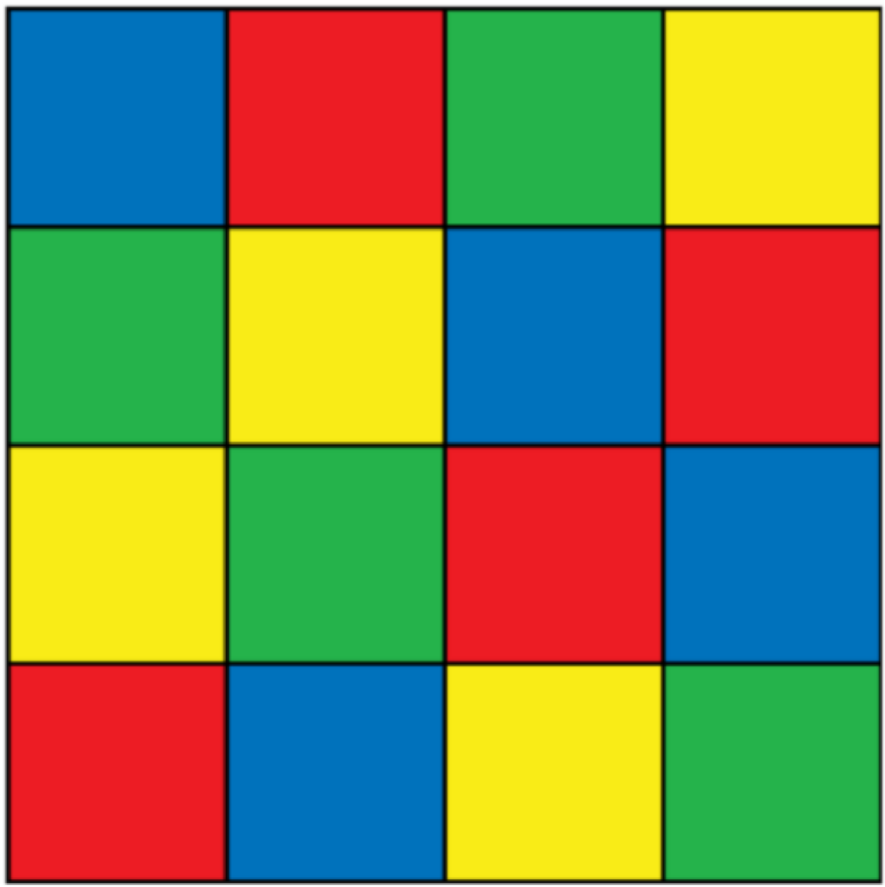

Can you please describe this image? In particular, please list the colors in different locations.

**ChatGPT**

The image shows a grid composed of smaller square blocks, each colored in either blue, red, green, or yellow. Starting from the top-left corner and moving row by row to the bottom-right corner, the colors are arranged as follows:

- Top row: Blue, Red, Green
- Middle row: Green, Yellow, Blue
- Bottom row: Yellow, Red, Blue

Each color appears to be solidly filled in its respective square, with no gradients or patterns.

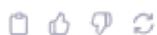

Figure S6

You

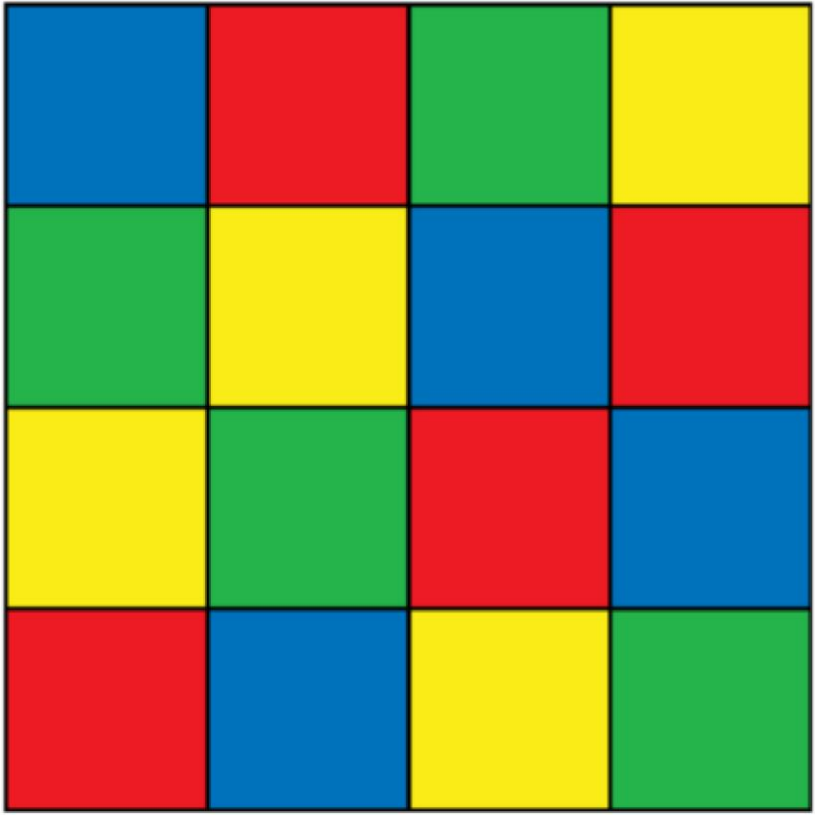

Can you please describe this image? In particular, please list the colors in different locations.

ChatGPT

The image appears to be a simple grid composed of 4×4 squares, each square filled with one of four colors: red, yellow, green, or blue. Starting from the top-left corner and moving to the right and then downwards, the colors are arranged as follows:

- Row 1: Blue, Red, Green, Yellow
- Row 2: Green, Yellow, Blue, Red
- Row 3: Red, Green, Yellow, Blue
- Row 4: Yellow, Blue, Red, Green

Each color is used exactly once in every row and column, similar to the rules of a Latin square.

‹ 2/2 ›

Figure S7

**You**

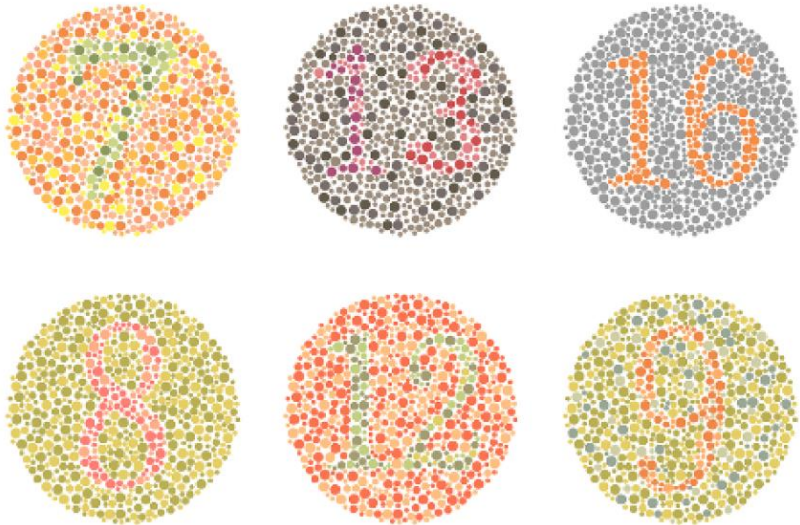

Can you please tell the numbers in this image?

**ChatGPT**

The image you've uploaded appears to be a set of Ishihara plates, which are used to test for color vision deficiencies. Here are the numbers I can identify within the circles:

1. The top-left circle appears to have the number '74'.
2. The top-middle circle seems to have the number '6'.
3. The top-right circle appears to have the number '16'.
4. The bottom-left circle seems to have the number '8'.
5. The bottom-middle circle appears to have the number '29'.
6. The bottom-right circle seems to have the number '45'.

These plates are designed in such a way that a person with normal color vision will see a certain number, while a person with a color vision deficiency may see a different number or no number at all. If you or someone else has difficulty seeing these numbers, it may be advisable to consult with an eye care professional for a complete evaluation.

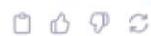

Figure S8